\documentclass[review,12pt]{elsarticle}

\usepackage{amsmath}
\usepackage{amsfonts}
\usepackage{lmodern,bm}   
\usepackage{float}               
\usepackage[T1]{sansmath} 
\usepackage{color}
\usepackage{url}

\newcommand{\lam}{\lambda}

\newcommand{\uv}{{\bf u}}

\journal{{\scriptsize Journal of Quantitative Spectroscopy \& Radiative Transfer}}

\makeatletter
\def\ps@pprintTitle{%
\let\@oddhead\@empty
\let\@evenhead\@empty
\def\@oddfoot{}%
\let\@evenfoot\@oddfoot}
\makeatother

\biboptions{numbers,sort&compress}

\usepackage[colorlinks=true,
allcolors=blue,
linktocpage=true,
bookmarks=true,
bookmarksnumbered=true,
bookmarksopen=true,
hyperindex,
breaklinks=true, 
pdfstartpage=1,%
pdfcenterwindow=false,%
pdffitwindow=true,%
pdfstartview={FitV},%
bookmarks=true,%
pdftitle={Thermal emission from a single glass fiber},
pdfauthor={Houssem Kallel, Joris Doumouro, Valentina Krachmalnicoff, Yannick De Wilde, Karl Joulain},
pdfcreator={pdfLaTeX with hyperref package},
pdfproducer={pdfLateX},
pdfdisplaydoctitle=true,
]{hyperref} 

\usepackage{lineno}

\begin{document}
\begin{frontmatter}

\title{Thermal emission from a single glass fiber}

\author[langevin,pprime]{Houssem Kallel\corref{hkallel}}
\ead{houssem.kallel@univ-poitiers.fr}

\author[langevin]{Joris Doumouro}
\author[langevin]{Valentina Krachmalnicoff}
\author[langevin]{Yannick De Wilde}
\ead{yannick.dewilde@espci.fr}
\author[pprime]{Karl Joulain}
\ead{karl.joulain@univ-poitiers.fr}

\address[langevin]{ESPCI Paris, PSL University, CNRS, Institut Langevin, 1 rue Jussieu, F-75005 Paris, France}
\address[pprime]{Institut Pprime, CNRS, Universit\'{e} de Poitiers, ISAE-ENSMA, F-86962 Futuroscope Chasseneuil, France}

\cortext[hkallel]{Corresponding author.}

\begin{abstract}
In this article, we study the thermal light emission from individual fibers of an industrial glass material, which are elementary building blocks of glass wool boards used for thermal insulation. Thermal emission spectra of single fibers of various diameters partially suspended on air are measured in the far field by means of infrared spatial modulation spectroscopy. These experimental spectra are compared with the theoretical absorption efficiency spectra of cylindrical shaped fibers calculated analytically in the framework of Mie theory taking as an input the measured permittivity of the industrial glass material. An excellent qualitative agreement is found between the measured thermal radiation spectra and the theoretical absorption efficiency spectra.
\end{abstract}



\end{frontmatter}


\section{Introduction}
Glass wool is a random assembly of glass fibers which has been used for decades in the thermal insulation of buildings. In conventional thermal insulation, the voids between glass fibers are filled with air. Since low fiber volume fraction is used (less than 10\%), the heat transfer is dominated by heat conduction through the air-filled regions, while heat convection is minimized. Typical apparent (effective) thermal conductivity of glass wool in air is about 40 mW/(m.K)~\cite{alam_VIPS_2011} which is close to the thermal conductivity of air (26 mW/(m.K)) \cite{strnad_stefan_1984}. Thick glass wool board of about 30 cm must be used to efficiently reduce the heat transfer~\cite{alam_VIPS_2011}. This conventional insulation is space and material consuming which does not fulfill the constraints imposed on the construction of future buildings. Substantially better thermal insulation is achieved when the air is evacuated from the solid fibrous assembly by lowering the air pressure to less than 0.1 mbar (10 Pa). The apparent thermal conductivity of a Vacuum Insulation Panel (VIP) with glass fibers as core material is expected to reach a value of few mW/(m.K)~\cite{kim_vacuum_2013}. In such a highly evacuated fibrous glass assembly, the gaseous thermal conductivity is suppressed and the heat transfer occurs exclusively by radiation and solid conduction. Heat conduction through a solid fibrous skeleton has been evaluated by Kwon et al. using simplified theoretical model \cite{kwon_effective_2009} and recently studied by Kallel et al.~\cite{kallel_computer_2019} by solving numerically the Laplace equation for three-dimensional complex geometries complementing earlier numerical works~\cite{arambakam_simple_2013,altendorf_influence_2014,huang_3d_2017}. Heat transfer by radiation is often estimated by considering a conduction model~\cite{tong_analytical_1980}~(also referred to as the Rosseland approximation). It is worth noting that the Rosseland analytical expression of the radiative conductivity of a gray optically thick medium \cite{modest_radiative_2013} is used by default for a fibrous glass assembly\cite{alam_VIPS_2011,kim_vacuum_2013}. In most
of the works reported to date, the radiative heat transfer through a fibrous assembly or its radiative properties are investigated by solving the radiative transfer equation (RTE)\cite{randrianalisoa_radiative_2017,arambakam_dual_2013,langlais_influence_1995,nicolau_spectral_1994,lee_effect_1989,lee_radiative_1986}.  In this work, we will focus on the thermal radiation from a single glass fiber. We compare theoretical findings with the measurement of the infrared (IR) thermal radiation from single micro-sized wires of various diameters from 4 to 9.5 $\mu$m which are compatible for use in VIPs~\cite{di_optimization_2013}. With respect to earlier reported optical measurements~\cite{schuller_optical_2009}, ours are performed with significantly lower temperature heating and done over an area where the object is totally surrounded by air. The lower temperature heating leads to higher quality Q factor of the emission while the suspended structure would avoid not only the emission signal from the substrate but also the influence of the substrate on the emission of the fiber. Recently, a novel thermometry platform based on the thermal fin model has been developed by Shin et al.~\cite{shin_far_field_2019} which enables the measurement of the emissivity of an individual suspended SiO2 nanoribbon at low temperature (down to 150 K). 

The spectral thermal radiance (or intensity) of an object ~\cite{kats_vanadium_2013} is equal to that of a blackbody $I_b(\lam,T)$ at wavelength $\lam$ and temperature $T$ (known as the Planck distribution function) multiplied by the emissivity of the object. When the object is opaque and its characteristic length is much larger than the thermal wavelength $\lam_{\mathrm{th}}$ (i.e. in the geometrical-optics approximation), thermal radiation is easily quantifiable by means of radiometry theory. The emissivity of such an object is function only of the Fresnel reflection coefficients \cite{joulain_surface_2005}. For an object with a characteristic length comparable to the thermal wavelength $\lam_{\mathrm{th}}$, thermal radiation can be determined using fluctuational electrodynamics \cite{rytov_principles_1989,joulain_surface_2005,volokitin_near-field_2007} which is based both on Maxwell equations and the fluctuation-dissipation theorem. Based on the theory of fluctuational electrodynamics, Golyk et al. \cite{golyk_heat_2012} have derived the heat radiation of an infinitely long cylindrical object. An easier way to evaluate the thermal radiation is 
to rely on the Kirchhoff's law \cite{kirchhoff_i._1860}. This law, which arises principally from reciprocity~\cite{li_nanophotonic_2018}, states that the emissivity of any object is equal to its absorptivity for every direction, wavelength and polarization. Originally, the Kirchhoff's law was derived in the geometrical optics approximation but it is valid even for small objects~\cite{miller_universal_2017}. The local form of this law, which was established for any finite-size object, equates the absorption cross-section density with the emissivity density \cite{rytov_principles_1989,greffet_light_2018}.
Bohren and Huffman demonstrated that a sufficient condition is required to satisfy the energy balance for any arbitrary isolated spherical particle~\cite{bohren_absorption_2007_sec4.7}. This condition imposes that the absorption efficiency, defined as the ratio of the absorption cross section to the geometric cross section, is equal to the emissivity. 

The outline of the paper is as follows. In Sec.~\ref{section1_experimental_apparatus}, we present the experimental set-up employed for the spectral measurement of the thermal radiation from a single fiber and we give also details on the acquisition procedure and the experiment conditions. In Sec.~\ref{section2_Comparison}, we first show theoretically how the measured thermal radiation spectrum is directly proportional to the absorption efficiency spectrum, and then analyse the dependence of the absorption efficiency spectrum on the fiber diameter and compare the measured spectra with computed theoretical spectra. Finally, in Sec.~\ref{section3_Conclusions}, we briefly summarize and draw some conclusions, and suggest some possible directions for future research.

\section{Description of the experimental apparatus and measurements}
\label{section1_experimental_apparatus}

The measurements have been realized with a new technique of IR spatial modulation spectroscopy (SMS), inspired from SMS technique previously developed for measuring the optical extinction cross-section of single nano-objects~using a laser~\cite{arbouet_direct_2004} or a white lamp~\cite{billaud_absolute_2010}. Details of the IR-SMS technique and of our experimental setup can be found in Ref.~\cite{li_near-field_2018}. It has recently proven its ability to detect the far-field thermal emission of a single subwavelength object~\cite{li_near-field_2018} and to spectrally analyze it with a Fourier transform IR (FTIR) spectrometer. The goal of the experiment is to extract the weak signal of the thermal emission from an individual object that coexists with a strong broadband background thermal emission. To achieve this goal, we spatially modulate the position of the studied object and use a synchronous detection processed by a lock-in amplifier. This allows us to record, at the modulation frequency, the thermal emission of the studied object with strongly reduced background contribution \cite{li_near-field_2018}. Without using spatial modulation, the studied object must be kept at relatively high temperature as in the case of 1-$\mu$m-diameter single silicon carbide (SiC) rod ($T$= 718 K) \cite{schuller_optical_2009}, this despite the fact that the SiC is highly efficient IR thermal emitter material.

\begin{figure}[!h]
\centering
\includegraphics*[width=9cm]{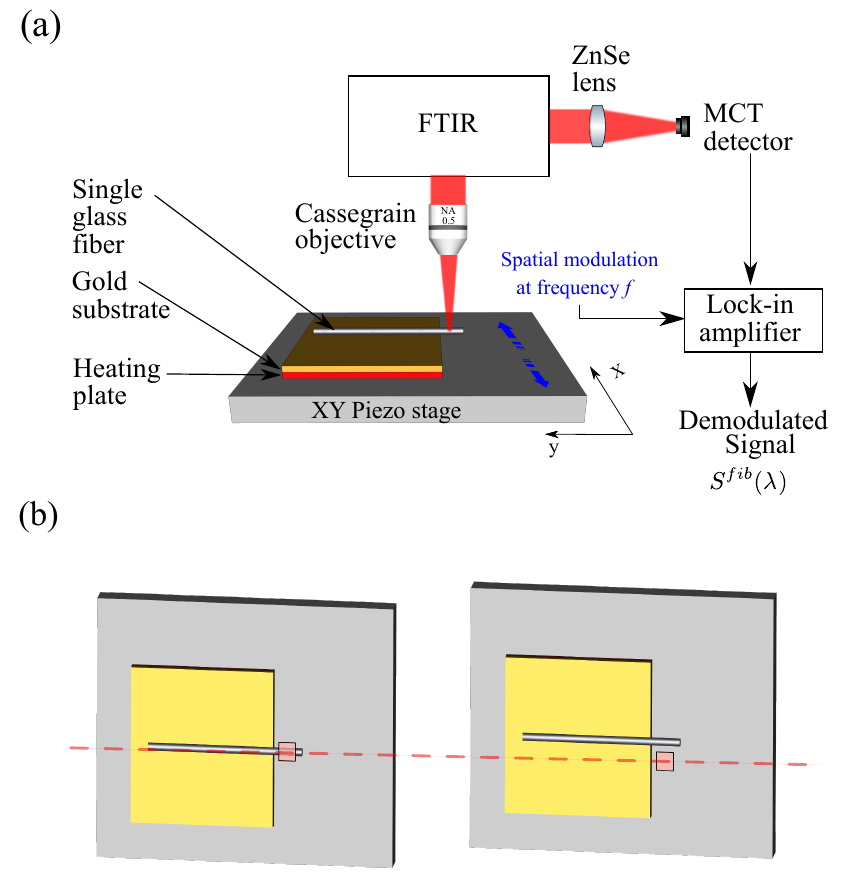}
\caption{(a) Schematic illustration of the spatial modulation spectroscopy setup for measuring the infrared thermal emission of an individual glass fiber. (b) Top view images showing the fiber at two different positions during the spatial modulation: inside (left) and outside (right) the 30 $\mu$m$\times$30 $\mu$m area probed by the setup.}
\label{fig1_exp_setup}
\end{figure}

The experimental set-up is shown schematically in Fig. \ref{fig1_exp_setup}. A single glass fiber is deposited on a gold plated silicon substrate and is kept in the object focal plane of a Cassegrain objective with numerical aperture (NA) of 0.5. The substrate is located on a heating plate. The edge of the heating plate is maintained at a constant temperature of 433 K. The blackbody used in our experiment is a black soot deposit whose emissivity is about 0.96 in the mid-infrared range~\cite{kats_vanadium_2013}. A piezoelectric XY stage holds the heating plate and allows to modulate the lateral position of the ensemble (heating plate, gold coated substrate and fiber) with a sinusoidal waveform. The spatial modulation is applied along the $x$ direction at frequency $f$=21 Hz with a peak-to-peak amplitude of 80 $\mu$m. The light collected by the Cassegrain objective passes first through a Fourier Transform IR (FTIR) spectrometer and is then focused by means of a zinc selenide (ZnSe) lens on a liquid nitrogen-cooled mercury-cadmium-telluride (MCT) IR detector. The magnification of the system is such that the IR detector captures light from a 30 $\mu$m$\times$30 $\mu$m area within the object focal plane of the Cassegrain objective. Initially, the XY stage is moved in a manner such that this detection area includes the fiber but is not lying on the substrate. However, care has been taken to cover the latter with gold (Au), i.e. a low emissivity material, as its radiation on the fiber could potentially produce a background scattering contribution to the detected signal. A lock-in amplifier is used to extract the amplitude of the first Fourier component (at the frequency $f$) from the output signal of the IR detector, by taking the sinusoidal drive signal as the lock-in reference signal. The extracted amplitude values are not only proportional to the thermal radiation emitted by the fiber but also to the response of the different components of the experimental setup. In order to eliminate the influence of the experimental setup, the spectrum recorded with a fiber sample $S^{fib}$ ($\lambda$) has been divided by the spectrum obtained for a blackbody reference heated at the same temperature of 433 K. We note that the measurements have been carried out in a specific spectral range, between 6 and 13 $\mu$m, which is directly related to the spectral response of the IR detector. Before performing the measurement, a glass fiber is locally heated by using an IR laser and pulled in order to decrease its diameter. The local diameter of the fiber is measured from bright-field optical images using a visible light illuminator. It is essentially constant in the whole spatial region that can be probed by the apparatus, which is 30 $\mu$m wide.

\section{Comparison experiment-theory}
\label{section2_Comparison}
\begin{figure}[!h]
\centering
\includegraphics*[width=7cm]{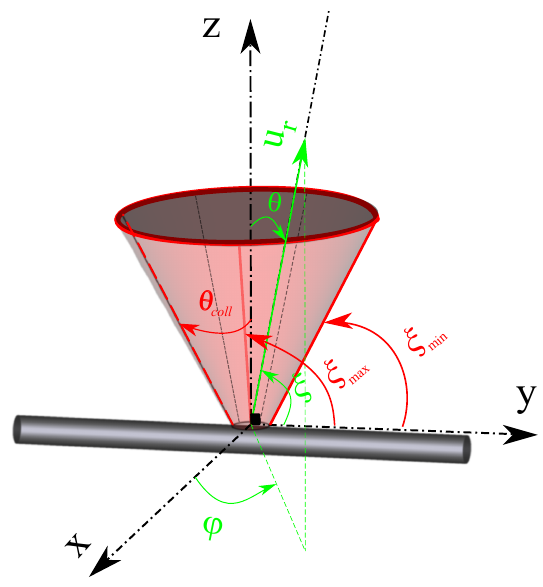}
\caption{Scheme showing the cone of half angle $\theta_{\mathrm coll}$ through which a portion of the light emitted by the fiber is collected by the objective.    
$\xi_{\mathrm min}$ (= 90$^\circ$ - $\theta_{\mathrm coll}$) and  $\xi_{\mathrm max}$ (= 90$^\circ$) are respectively the minimum and maximum
possible values of the angle of incidence $\xi$.}
\label{fig2_Emission_collect}
\end{figure}

Let us recall first how the thermal emission from a single isolated fiber can be determined theoretically. The power radiated per unit wavelength interval by a fiber in an elementary solid angle $d\Omega= sin{\theta}d{\theta}d\varphi$ around the direction $\uv_r$ at a given wavelength $\lam$ can be expressed by~\cite{greffet_light_2018}

\begin{equation}
\label{emission}
dP_{e}(\lam)=~\sigma_{\mathrm{abs}}(-\uv_r,\lam)I_b(\lam,T)~d\Omega
\end{equation}
where $I_b(\lam,T)$ is the spectral intensity of a blackbody radiation and $\sigma_{\mathrm{abs}}(-\uv_r,\lam)$ is the absorption cross section for non-polarized incident light in the direction -$\uv_r$ with a wavelength $\lam$ (see Fig.~\ref{fig2_Emission_collect}). As mentioned in the introduction, the absorption cross-section $Q_{\mathrm{abs}}$ normalized with respect to the geometrical cross section of the fiber ($\sigma_{\mathrm{geo}}= D~L$, $D$ and $L$ are the diameter and the length of the fiber respectively), also known as the absorption efficiency, is equal to the emissivity of the fiber~\cite{bohren_absorption_2007_sec4.7}. This equality is nothing other than the Kirchhoff's law \cite{kirchhoff_i._1860} for a finite-size object for every direction, wavelength and polarization.

Integration of Eq.\eqref{emission} over the entire far-field sphere and over all wavelengths results in total thermal radiation emitted by the isolated fiber~\cite{bohren_absorption_2007_sec4.7} 
\begin{equation}
\label{ }
P_{tot}=~4\pi \int_0^\infty I_b(\lam,T)~Q_{\mathrm{abs}}(\lam)~\sigma_{\mathrm geo}~d\lam.
\end{equation}
We assumed that the normalized absorption cross section $Q_{\mathrm{abs}}(\lam)$ is the same whatever the direction of the incident light.

As shown in Fig.\ref{fig2_Emission_collect}, the amount of thermal radiation emitted by the fiber and collected by the apparatus, depends on the collection angle of the objective $\theta_{\mathrm coll}$ (=arcsin(${\mathrm {NA}}$)), and on the geometrical cross-section probed by the apparatus $\sigma_{\mathrm {probed}}=D~l$, where l=30 $\mu$m is the size of the detection area optically conjugated with the IR detector,

\begin{equation}
\label{eq1}
P_{\mathrm coll}\propto \int_{6{\mu}m}^{13{\mu}m} d\lam\int_0^{2\pi}d\varphi\int_0^{\theta_{\mathrm coll}}~d\theta~sin\theta~Q_{\mathrm{abs}}(\theta,\varphi,\lam)I_b(\lam,T)D~l.
\end{equation}
The thermal radiation spectrum of the fiber $S^{\mathrm fib}(\lam)$ can be roughly expressed as follows

\begin{equation}
\label{eq2}
S^{\mathrm fib}(\lam)\propto \int_{\scriptstyle {90{^\circ} - \theta_{\mathrm coll}}}^{90^\circ}~d\xi~Q_{\mathrm{abs}}(\xi,\lam)~I_b(\lam,T) D~l
\end{equation}
where $\xi$ is the angle of incidence. One can normalize the spectrum $S^{\mathrm fib}(\lam)$ by dividing it with the thermal radiation spectrum of a blackbody and thus obtain

\begin{equation}
\label{eq3}
S_{\mathrm N}^{\mathrm fib}(\lam)\propto  \int_{{90{^\circ} - \theta_{\mathrm coll}}}^{90^\circ}~d\xi~Q_{\mathrm{abs}}(\xi,\lam) D.
\end{equation}

Since the numerical aperture of the objective used in the experimental setup is equal to 0.5 (see Fig.~\ref{fig2_Emission_collect}), the angle of collection $\theta_{\mathrm coll}$ is 30${^\circ}$ and the angle of incidence $\xi$ varies from 60 to 90${^\circ}$. Over this incidence angle range, the absorption efficiency spectrum is approximately the same and does not differ significantly from that obtained for $\xi$= 90${^\circ}$ (data not shown). This latter result, reported already for other wires~\cite{kallel_tunable_2012}, allows us to write the following estimation  

\begin{equation}
\label{eq_thermal_radiation}
S_{\mathrm N}^{\mathrm fib}(\lam)\propto Q_{\mathrm{abs}}(\xi= 90^\circ,\lam) D
\end{equation}

\begin{figure*}[ht!]
\centering
\includegraphics*[width = 0.5\textwidth]{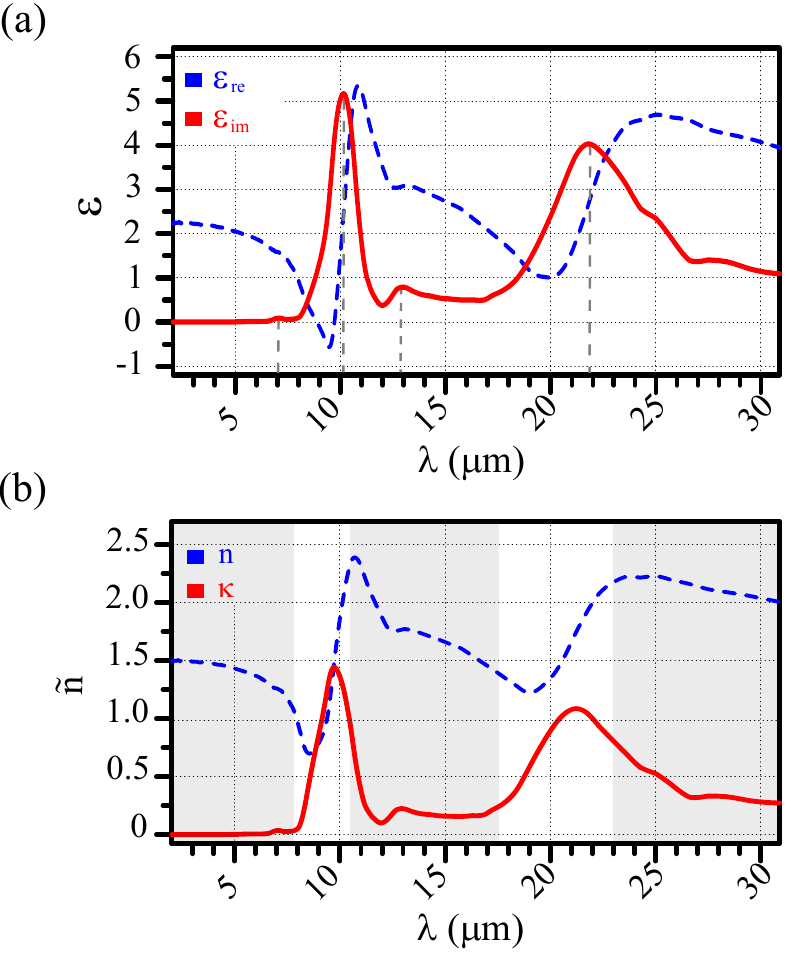}
\caption{(a) Complex permittivity $\varepsilon$ = $\varepsilon_{\textrm{re}}$~+~i~$\varepsilon_{\textrm{im}}$ and (b) complex refractive index $\tilde{n}$ = $n$~+~i~$\kappa$ of the bulk industrial glass material. The vertical dashed lines indicates in (a) the characteristic peaks of the spectrum $\varepsilon_{\textrm{im}}$($\lam$). The grey coloured regions shown in (b) correspond to wavelength regions in which the refractive index $n$ is relatively higher than the extinction coefficient (or extinction index) $\kappa$.}
\label{fig3_CM210_perm_indx}
\end{figure*}

The absorption efficiency $Q_{\mathrm{abs}}$ for non-polarized light illumination (as well as the scattering efficiency $Q_{\mathrm{sca}}$ ) can be calculated using the Mie theory ~\cite{bohren_absorption_2007} which solves the problem of scattering of a plane wave by an infinitely long cylinder at oblique incidence. The non-polarized light illumination is the average of two orthogonal polarized incident lights: Transverse electric (TE) and transverse magnetic (TM), which implies that the dimensionless far-field quantity $Q_{\mathrm{abs,sca}}(\xi,\lam)=\frac{Q_{\mathrm{abs,sca}}^{\mathrm{TE}}(\xi,\lam) + Q_{\mathrm{abs,sca}}^{\mathrm{TM}}(\xi,\lam)}{2}$. For details, the reader can refer to \cite[][chapter 8, section 8.4]{bohren_absorption_2007} and \cite[][Appendix, section 1]{kallel_tunable_2012}. The analytical Mie theory has the advantage to allow an exhaustive study. The absorption/elastic scattering spectra of individual wires, which are calculated using this theory, are often compared with those measured qualitatively and not quantitatively\cite{cao_engineering_2009,bronstrup_Optical_2010,kallel_tunable_2012}. The quantitative comparison requires, in addition to the accurate measurement through a rigorous calibration, the use of a numerical approach such as the finite element method (FEM) or the boundary element method (BEM).

The material constituting the fiber is an industrial glass. Its complex permittivity and refractive index, determined at Saint-Gobain Recherche \cite{langlais_influence_1995}, are plotted as a function of the incident light wavelength in Figs.~\ref{fig3_CM210_perm_indx}(a) and \ref{fig3_CM210_perm_indx}(b), respectively. These bulk values are considered to be the same for the glass fiber. Note that the temperature and strain effects on the complex permittivity are assumed to be negligible. In addition, the glass fiber is assumed to be surrounded by vacuum.
\begin{figure*}[ht!]
\centering
\includegraphics*[width = 0.8\textwidth]{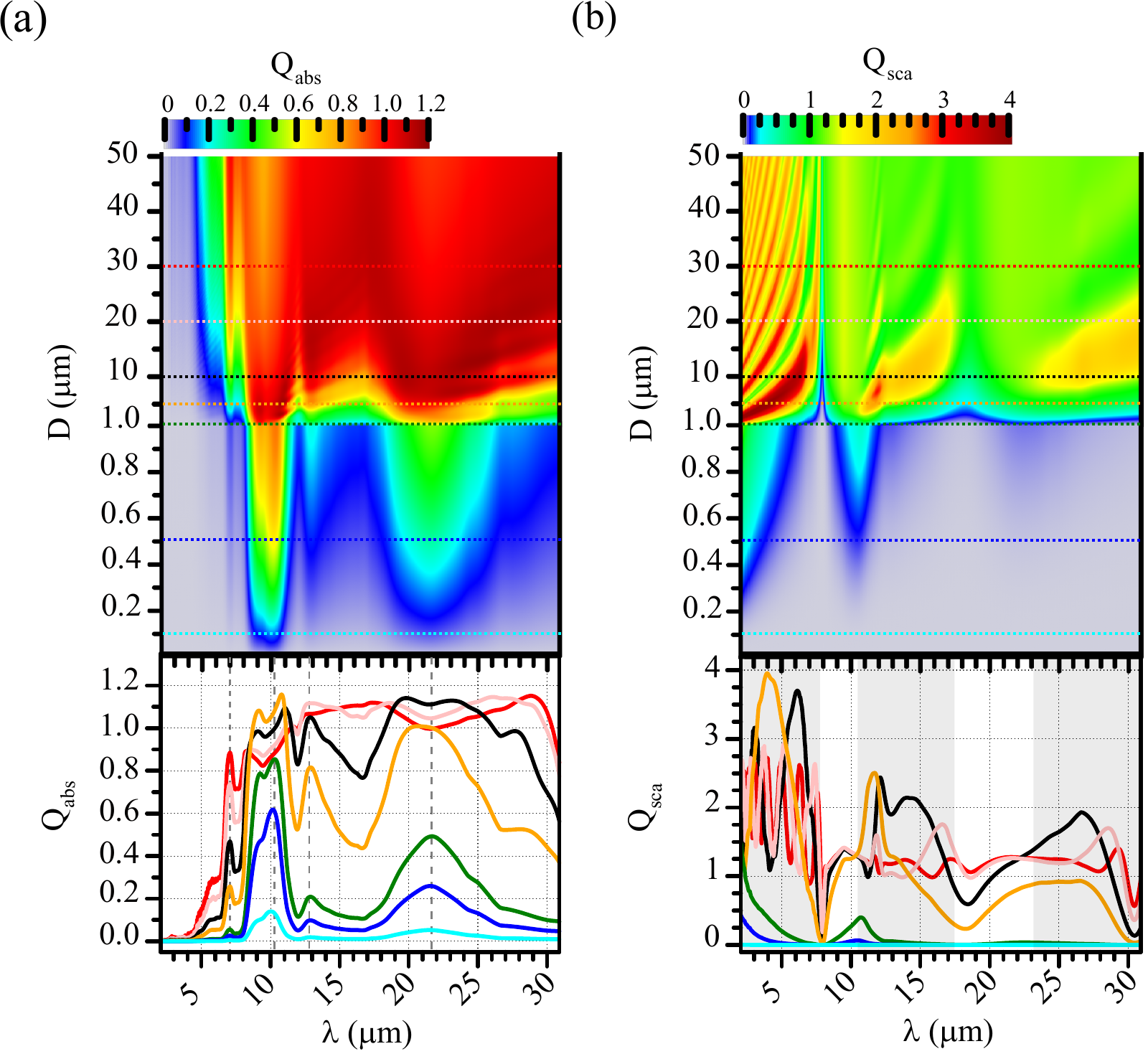}
\caption{(a) Absorption efficiency $Q_{\textrm{abs}}$ and (b) scattering efficiency $Q_{\textrm{sca}}$ of a glass fiber in vacuum as a function of its diameter \textit{D} and the incident light wavelength $\lambda$ under normal incidence illumination ($\xi$= 90$^{\circ}$) and for non-polarized incident light. The $Q_{\textrm{abs}}$ spectra and $Q_{\textrm{sca}}$ spectra are taken for a glass fiber with selected diameters : $D$ = 0.1, 0.5, 1, 5, 10, 20 and 30 $\mu$m. Dashed horizontal lines on the map with various colors correspond to the spectra shown.}.
\label{fig4_Qabs_sca_Glass}
\end{figure*}

The computed absorption efficiency spectra are shown in a color map as a function of the fiber diameter in the upper part of Fig.~\ref{fig4_Qabs_sca_Glass}(a). Note that the efficiency values are coded with different colors. The spectra of fibers with selected diameters ($D$ = 0.1, 0.5, 1, 5, 10, 20 and 30 $\mu$m) are plotted in the bottom part of Fig.~\ref{fig4_Qabs_sca_Glass}(a). Results are presented for an incident light with wavevector normal to fiber axis (normal incidence), corresponding to an angle of incidence $\xi= 90^\circ$.

We initially focus on the absorption of small glass fibers with diameter below than or equal to 1 $\mu$m. In the color map in Fig~\ref{fig4_Qabs_sca_Glass}(a), we can distinguish two relatively broad absorption bands: the first one, extending approximately from 8 to 12 $\mu$m, the other one, from 18 to 26 $\mu$m. They are centred around 10 $\mu$m and 22 $\mu$m, respectively. An additional weaker absorption band appears near 13 $\mu$m and a very low intensity and tiny peak at 
7 $\mu$m. We can observe that the absorption increases over the entire spectrum when the fiber diameter increases. We remark that the spectrum of the absorption efficiency of small glass fibers is similar in shape to the spectrum of the imaginary part of the permittivity $\varepsilon_{\textrm{im}}$($\lam$) (see the solid red curve in Fig.~\ref{fig3_CM210_perm_indx}(a)). The peaks observed in the absorption spectra (see the solid cyan, blue and green curves in Fig.~\ref{fig4_Qabs_sca_Glass}(a)) are indicated by vertical dashed lines. Their spectral positions coincide with the positions of the peaks in the spectrum of the imaginary part of the permittivity (indicated also by vertical dashed lines).   The absorption bands are absolutely identifiable in the extinction coefficient spectrum $\kappa$($\lam$) (see the solid red curve in Fig.~\ref{fig3_CM210_perm_indx}(b)). The three main absorption bands are attributed to the resonance of Si-O-Si vibrations \cite{kitamura_optical_2007}. It is worth pointing out that the industrial glass is composed mainly of silica, SiO$_{2}$. This explains why the complex dielectric function and complex refractive index of the industrial glass, as shown in Fig. \ref{fig3_CM210_perm_indx}, are so similar to those of silica glass \cite{kitamura_optical_2007}. The weak peak at 7 $\mu$m is owing to the presence of boron oxide (B$_{2}$O$_{3}$) in the composition of the industrial glass~\cite{langlais_influence_1995}. Thus, we can deduce that the absorption of the glass fiber is strongly dependent on the absorption properties of the industrial glass material. Moreover, the absorption band, located between 8 to 12 $\mu$m, contains a peak around 9 $\mu$m. This peak appears in the spectral region where the real part of the permittivity $\varepsilon_{\textrm{re}}$ becomes negative. Ideally, the surface phonon-polariton resonance (SPhPR) takes place at the wavelength when the complex permittivity $\varepsilon$ is purely real and equals to -1.

We now turn our attention to the absorption of glass fibers with diameter above  1 $\mu$m. As the diameter increases, the absorption band at longer wavelengths  becomes wider and wider. Furthermore, in the spectral range between this band and the neighbouring band around 13 $\mu$m, the glass fiber absorbs more and more efficiently with increasing the diameter. Since the fiber is constituted by a low refractive index material (see blue curve in Fig. 3 (b)), the increased absorption cannot be attributed to whispering Gallery modes due to the inefficient light trapping caused by the small index contrast between the fiber and the surrounding medium (i.e. vacuum). Thus, it is not related to an optical resonance effect but rather to the fact that the light attenuation becomes more and more important as the thickness of the object increases.  This behaviour may be described by the the Beer-Lambert law. It is, in fact, very similar to that of a glass slab of thickness $D$.  The absorption efficiency of a fiber is approximately equal to the absorbance of a slab, i.e. $Q_{\mathrm{abs}}{\approx}A = 1 - e^{-{\alpha}D}$ where ${\alpha}=\dfrac{4\pi\kappa}{\lambda}$ is the absorption coefficient. Thick fibers exhibit a nearly flat spectral absorption for wavelengths longer than 12.5 $\mu$m. For instance, the pink curve in the bottom panel of Fig.~\ref{fig4_Qabs_sca_Glass}(a) is the spectrum for a 20 $\mu$m diameter fiber that shows particularly absorption efficiency values close to 1.1 in the 12.5-29 $\mu$m wavelength range. It is interesting to note that, for diameters ranging from 15 to 50 $\mu$m, the absorption efficiency is higher than 0.9 and exceeds unity over a broad spectral range (see Fig.~\ref{fig4_Qabs_sca_Glass}(a)), top). This finding is consistent with the results reported by Golyk et al.~\cite{golyk_heat_2012} showing that for large diameters (i.e. D $\geq$ 20 $\mu$m) the total heat radiation of SiO$_{2}$ cylinder exceeds that of the plate. It should be mentioned that the short-wavelength absorption band intensity decreases slightly while the absorption peak at 7 $\mu$m increases in strength as the diameter of the fiber diameter increases.  The discussion above remains valid for glass spheres. In fact, the absorption efficiency spectrum of a SiO$_{2}$ sphere, as plotted in Fig.~1(b) of Ref.~\cite{fernandez_super_2018}, exhibits a diameter-dependence similar to that of a glass fiber, despite the discrepancies in shape and complex permittivity. 

\begin{figure*}[ht!]
\centering
\includegraphics*[width = 0.6\textwidth]{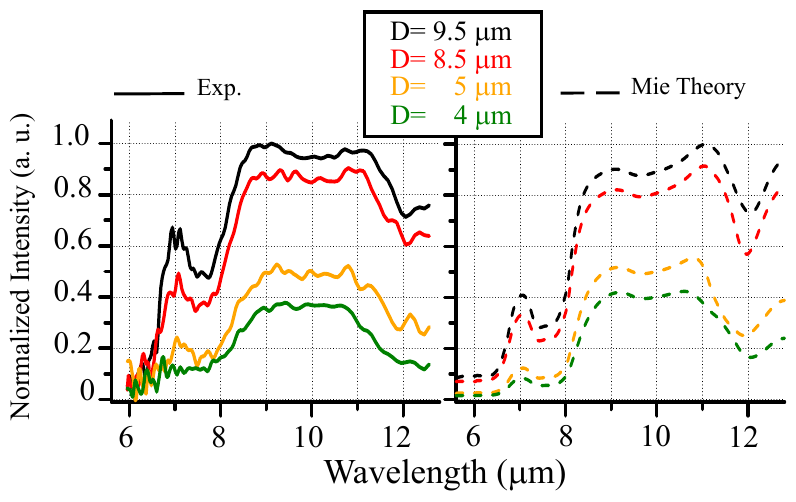}
\caption{Normalized measured thermal radiation spectra (left) and normalized absorption efficiency spectra multiplied by the fiber diameter, ie, $Q_{\mathrm{abs}}(\xi= 90^\circ,\lam){\times}D$ (right) for different fiber diameters. The normalization is done with respect to the maximum of the spectrum obtained for a diameter of 9.5 $\mu$m.}
\label{fig5_compar_results}
\end{figure*}

From Eq.\ref{eq_thermal_radiation}, one expects the measured thermal radiation to be proportional to the product of the absorption efficiency and the fiber diameter. We now compare in Fig.\ref{fig5_compar_results} the measured spectra with the absorption efficiency spectra calculated with the Mie theory, multiplied by the fiber diameter. Measurements have been performed using our experimental apparatus, as described in Section~\ref{section1_experimental_apparatus}, on single glass fibers of various diameters 4.0, 5.0, 8.5 and 9.5 $\mu$m. The wavelength range covered by the experimental spectrum extends   
from 6 to 13 $\mu$m. As indicated above, in this spectral region $Q_{\mathrm{abs}}$($\lam$) is expected to show only a small peak at 7 $\mu$m followed by a broad band between 8 and 12 $\mu$m. We can distinguish these two main spectral features in the experimental curves, as shown in the left panel of Fig. \ref{fig5_compar_results}. We also note a drop in intensity of the 7 $\mu$m absorption peak as the diameter shrinks, in agreement with theoretical expectations. In Fig.~\ref{fig4_Qabs_sca_Glass}(b) we plot the calculated scattering efficiencies, in the same manner as the absorption efficiencies in Fig.~\ref{fig4_Qabs_sca_Glass}(a). The $Q_{\mathrm{sca}}$ spectra are different from those of $Q_{\mathrm{abs}}$ whatever the fiber diameter. The light scattering is more efficient for thick fibers (ie, $D>$ 1 $\mu$m). For thick fibers, efficient light scattering occurs when the refractive index $n$ is much higher than the extinction coefficient $\kappa$ (see grey coloured regions in Fig.~\ref{fig3_CM210_perm_indx}(b) and lower panel of Fig.\ref{fig4_Qabs_sca_Glass}(b)). In the corresponding wavelength regions, we observe branches showing high efficiency values on the upper panel of Fig.\ref{fig4_Qabs_sca_Glass}(b). These branches are not associated with specific Mie resonances but are due to the overlap of different resonance modes (noting that the industrial glass is low-refractive-index material). The above discussion raises the question of whether the scattering of the background thermal radiation by the fiber contributes considerably to the signal measured. In the lower panel of Fig.~\ref{fig4_Qabs_sca_Glass}(b), the orange curve shows the spectrum $Q_{\mathrm{sca}}$($\lam$) for a diameter of 5 $\mu$m  while the black curve for a diameter of 10 $\mu$m. Enhanced scattering is observed in both spectra in the range of 6-6.5 $\mu$m. However, the measured signal in this spectral region is very weak for all of the diameters studied, as shown in Fig.\ref{fig5_compar_results}. Furthermore, for a fiber of 5 $\mu$m diameter the scattering spectrum contains a narrow peak around 11.6 $\mu$m and for a 10 $\mu$m diameter fiber a peak around 12.2 $\mu$m. These spectral features are totally absent in the experimental spectra. The distinct differences between scattering efficiency and experimental thermal radiation spectra indicate that the contribution of scattering of the background thermal radiation is unsignificant in the measured signal using the spatial modulation technique. Finally, it is interesting to mention that Fig.~\ref{fig5_compar_results} shows that the intensity of the whole thermal radiation spectrum increases much more than twofold while increasing the fiber diameter from 4 to 9.5 $\mu$m. Consequently, the total radiating power is expected at least to double and so the larger diameter would lead to higher radiation. To minimize the heat transfer by radiation, it would be necessary to use fibers with smaller diameters-this assuming of course a weak optical interaction between the fibers.

\section{Conclusions}
\label{section3_Conclusions}
We have measured the thermal radiation spectra of individual glass fibers using a method based on the modulation of the sample position. We find that there is an obvious similarity between the measured spectra and the absorption efficiency spectra calculated by means of Mie theory. The scattering of the background thermal radiation did not appear to contribute significantly to the measurement. The thermal radiation spectral features of a single glass fiber or equivalently the absorption, are strongly linked to the optical properties of the glass material. Since the glass wool used for thermal insulation is composed by an assembly of fibers, future research should focus on the radiative properties of glass fibrous assembly. We intend to calculate it using a boundary element method \cite{reid_efficient_2015,scuff_em_online_doc} which requires only the mesh of the surface that we have succeeded in creating \cite{kallel_computer_2019}.
 The fluctuating-surface-current formulation developed by Rodriguez et al. \cite{rodriguez_fluctuating-surface-current_2012,rodriguez_fluctuating-surface-current_2013,polimeridis_fluctuating_2015} has proven to be a promising approach for analysing the radiative heat transfer by arbitrary geometries.

\section*{Acknowledgements}
We thank Saint Gobain Recherche for its support, the industrial glass samples supply, and the industrial glass optical properties measurements. We also thank Claire Li (I. Langevin) for helping to perform the spectroscopy measurements on single fibers. This work received financial support from LABEX WIFI (Laboratory of Excellence within the French Program Investments for the Future) under references ANR-10-LABX-24 and ANR-10-IDEX-0001-02 PSL*, and Agence Nationale de la Recherche (ANR), Project CarISOVERRE under reference ANR-16-CE09-0012. The work also pertains to the LABEX INTERACTIFS under the reference ANR-11-LABX-0017- 01.

\bibliography{references} 







\end{document}